\documentstyle[aps,psfig,12pt]{revtex}

\begin{document}

\begin{center}

{\Large Damping of Oscillations in Layer-by-Layer Growth}
\vspace{1cm}

H.~Kallabis~$^{(*)}$, L.~Brendel~$^{(*)}$, J.~Krug~$^{(**)}$, D.~E.~Wolf~$^{(*)}$
\vspace{0.5cm}

{\small (*) HLRZ, Forschungszentrum J\"ulich, D-52425 J\"ulich,
  Germany\\and\\
FB 10, Gerhard -- Mercator -- Universit\"at, D-47048 Duisburg, Germany} 

{\small (**)
 Fachbereich Physik, Universit\"at GH Essen, D-45117 Essen, Germany} \\

\vspace{1cm}
{\bf Abstract}\\

\end{center}

We present a theory for the damping of layer-by-layer growth oscillations
in molecular beam epitaxy. The surface becomes rough on distances larger
than a {\em layer coherence length} which is substantially larger than
the diffusion length. The {\em damping time} can be calculated by a
comparison of the competing roughening and smoothening mechanisms.
The dependence on the growth conditions, temperature and deposition rate,
is characterized by a power law. The theoretical results are confirmed
by computer simulations.

\vfill

\section{Introduction}
\label{introduction}

Basic questions of crystal growth have been a source of 
inspiration in physics for many years
\cite{VP95}.
Several concepts of general theoretical importance were originally
developed in this area, but a good deal of the fascination is also due
to the technological relevance.

Here we want to address layer-by-layer growth or Frank-van der 
Merwe growth, as it is also called \cite{bauer}. This is a growth mode in 
molecular beam epitaxy which allows well controlled manipulation of
e.g. chemical composition in layers down to atomic thickness. It is
therefore particularly suited for the fabrication of electronic
devices. 

In the most elementary model of layer-by-layer growth 
atoms are deposited under high vacuum conditions onto 
a (100) or (111) surface, for instance. 
The adatoms diffuse on the surface until they meet 
one another to form dimers which then grow into islands of monoatomic
height. Their edges capture most of the adatoms during the
deposition of one monolayer. When the available island edges become
less due to coalescence, the formation of dimers and islands in the
next layer begins. Hence the density of atomic steps like all other
quantities which are sensitive to the surface morphology
oscillate. These oscillations are the real hallmark of layer-by-layer
growth. They can be used to monitor the film thickness easily.

In step flow growth of a vicinal surface, where the
nucleation of islands on the terraces can be neglected, as well as for
rough surfaces the oscillations are absent.
In practice, one cannot prepare substrates without a miscut. It may
happen to be very weak, but it cannot be controlled. 
Layer-by-layer growth with its oscillations can only be seen if some
terraces between steps are wide enough to allow for the formation of
islands. 

Generically, the oscillations are damped: Layer-by-layer growth is
only a transient. The reason can be either the emergence of a smooth
surface with step flow 
or of roughness \cite{evans}, two entirely different diagnoses for the same
symptom. However, increasing the temperature the damping should become
stronger in the first and weaker in the second case, allowing to
discriminate the two. Roughness itself can have two different reasons.
If interlayer transport is inhibited by step edge barriers, adatoms
accumulate on top of islands, and one obtains the growth instability
predicted by J. Villain \cite{villain91}, demonstrated and analysed in
computer simulations \cite{siegert} and observed in many
experiments \cite{experiments}. If no instability occurs, the surface may still
roughen due to fluctuations in the beam intensity  (shot noise).
This is the case which will be discussed below in more detail.

Layer-by-layer growth is also possible if one starts from a
patterned substrate rather than a perfectly smooth one, and one
may wish to optimize the growth conditions such that the pattern
stays intact for several layers. This optimization process is
nontrivial: On one hand, the islands nucleate with a characteristic
distance $l$, which sets the spatial resolution for the pattern in
subsequent layers. On the other hand, deposition, diffusion and island
nucleation are stochastic processes and lead to the accumulation of
errors in the pattern, the thicker the film grows. Often, reducing these errors
implies an increase of $l$ and hence a worse resolution of the pattern.

In order to understand layer-by-layer growth, one therefore has to
investigate spatial correlations, temporal correlations and finally
the emergence of surface roughness, which determines, for how many
layers the oscillations can be seen. 

The spatial
correlations in layer-by-layer growth manifest themselves through the
characteristic distance $l$ of nucleation events within a layer
mentioned above.
The dependence of $l$ on growth conditions has already been well
studied 
\cite{zinsmeister,stoyanov,venables,villain92,pimpinelli,wolf,pimpinelli96}, 
and the key
result needed later is 
\begin{equation}
l \propto (D/F)^{\gamma},
\label{l}
\end{equation}
where $D$ and $F$ are the surface diffusion constant and the
deposition rate, respectively. The exponent $\gamma$ depends on the 
adatom diffusion process and the (fractal) dimension of the
islands. It also depends on whether or not desorption of adatoms or
diffusion of dimers or larger clusters are negligible, as we assume in
the simulations presented below. Finally, $\gamma$ is a function of
the critical island size $i^*$, which is defined by the size $i^*+1$ of 
the smallest island, which is stable enough that it never decays
before capturing the next adatom. 

The temporal correlations are interesting, because they determine how
well a nucleation pattern is reproduced in subsequent layers. For
example there is a high probability that the first island nucleates
on top of the first one of the previous layer, because this is the
area where adatoms accumulate most likely. In fact, it was shown that
the autocorrelation function of the first nucleation events decays
rather slowly with time like $t^{-1/2}$ \cite{somfai}. Of course, this
power law is only valid as long as layer-by-layer growth persists. It
will be cut off by the time $\tilde t$, which characterizes the damping of
the oscillations.

The main subject of this paper is the emergence of surface roughness 
responsible for the damping of the oscillations. 
The simulation results show that layer-by-layer growth goes on forever
if the linear size of the system is smaller than a layer coherence 
length $\tilde
l$. Up to this length the layers grow coherently, for larger distances
they get out of phase. Remarkably, $\tilde l$ is much larger than the
characteristic distance between islands, $l$. 
The surface becomes rough on
scales larger than $\tilde l$ rather than the diffusion length $l$. 

In order to study kinetic roughening one may average the film thickness
over the distance $\tilde l$. Then one cannot resolve individual
islands any more, but still sees the dephasing between
layers. Phenomena on this scale can be described by
continuum equations, which provide the most transparent theoretical framework 
in which to discuss the smoothening mechanisms competing with the shot noise.
The layer coherence length $\tilde l$ as well as the damping time $\tilde t$
play an important r\^ole for kinetic roughening as natural cutoffs of the
continuum growth equation at small length and time scales. This idea will be
worked out in section \ref{theory}.

Recently, it has been 
discovered \cite{brendel,wolf} that $\tilde t$ depends on the growth conditions
like
\begin{equation}
F\tilde t \propto (D/F)^{\delta} .
\label{tilde_t}
\end{equation}
This law will be explained in section \ref{exponent} based on the results of
section \ref{theory}.
Both sections, \ref{theory} and \ref{exponent}
rely on dimensional arguments. In section \ref{current} the key results
of the preceding section are rederived avoiding dimensional arguments.
Section \ref{competition} contains a discussion of crossovers due to
different smoothening mechanisms, and section 
\ref{numerics} reviews the numerical results confirming the theoretical 
picture.

\section{A theory for the layer coherence length and the damping time}
\label{theory}

The transition from layer-by-layer growth with its oscillations to kinetic
roughening happens at time $\tilde t$, when the film thickness varies over
the distance $\tilde l$ by about one atomic layer.
For $t>\tilde t$ one expects that the surface shows
self affine scaling \cite{family_vicsek}:
\begin{equation}
w(t) \approx a_{\perp}(\xi(t)/\tilde l)^{\zeta} \qquad {\rm with} \qquad
\xi(t) \approx \tilde l \ (t/\tilde t)^{1/z} .
\label{w_scaling}
\end{equation}
Here, $w$ is the root mean square variation of the film thickness, $a_{\perp}$ 
the thickness of one atomic layer, and 
$\xi$ the correlation length up to which the surface 
roughness has fully developed until time $t$. $\zeta$ is the roughness
exponent and $z$ the dynamical exponent.

$\tilde t$ is the time at which a 
continuum description of kinetic roughening becomes appropriate.
Whenever desorption and the formation of defects in the 
growing film can be neglected the equation of motion must have the form
\begin{equation}
\partial_t h=- \nabla \cdot{}{\bf j} + \eta ,
\label{cKPZ}
\end{equation}
where $h$ is the deviation of the 
film thickness from its average value and $\eta({\bf x},t)$ 
denotes the shot noise
with correlator
\begin{equation}
\langle \eta({\bf x},t) \eta({\bf x'},t') \rangle = {\cal F}\delta^d({\bf x}-{\bf x'}) \delta(t-t') .
\label{noise_correlator}
\end{equation}
$d$ is the surface dimension.
In the conserved KPZ (cKPZ) equation, which has been widely discussed 
\cite{villain91,wolf_villain,dasSarma,CKPZ}
in the context of molecular beam epitaxy,
the adatom current has two terms, one driven by differences in 
the surface curvature and the second one by differences in the squared
surface tilt:
\begin{equation}
{\bf j} = \nabla [K \nabla^2 h + \lambda (\nabla h)^2].
\label{surface_current}
\end{equation}

Eq.(\ref{w_scaling}) shows that the only characteristic length, time and 
height entering the description of the rough surface (coarse grained on 
scale $\tilde l$) are $\tilde l$, $\tilde t$ and $a_{\perp}$, respectively.
Therefore $K$, $\lambda$ and ${\cal F}$ must be functions of these three
quantities. For example, $\lambda$ has the dimension length$^4$ height$^{-1}$ 
time$^{-1}$. This implies that it must be the product of a dimensionless factor
and $\tilde l^4/a_{\perp}\tilde t$. Similarly one obtains
\begin{equation}
K \approx a_{\perp}\lambda \approx \tilde l^4/\tilde t
\label{K1}
\end{equation}
According to (\ref{cKPZ}), $\eta$ has the dimension height/time.
Taking the dimensions of the $\delta$-functions in (\ref{noise_correlator})
(length$^{-d}$ and time$^{-1}$, respectively) into account, one finds that
${\cal F}$ is
\begin{equation}
{\cal F} \approx a_{\perp}^2 \tilde l^d/ \tilde t,
\label{F1}
\end{equation}
up to a dimensionless factor.

\section{Connection with submonolayer physics}
\label{exponent}

In order to derive (\ref{tilde_t}) from (\ref{K1}), (\ref{F1}) one has to know, how
$K$ (or $a_{\perp}\lambda$) and ${\cal F}$ depend on $D$ and $F$. 
This question will be answered in the following.

The physics of kinetic roughening should be determined by the same
microscopic processes that are also responsible for the phenomena in the
submonolayer regime. There, the characteristic time is the layer completion
time, 
\begin{equation}
\tau = (F a^d)^{-1},
\label{tau}
\end{equation}
and there are two characteristic lengths,
the diffusion length, $l$, and the lattice constant along the surface, $a$. 
Therefore, it must be possible to express 
$K$, $\lambda$ and ${\cal F}$ in terms of $l$, $a$, $\tau$ and $a_{\perp}$.

The coefficients $\lambda$ 
and $K$ characterize the morphology dependence of the nonequilibrium 
adatom density, which drives the surface current (see Section \ref{current}).
The most important morphological feature is the typical distance
between islands. 
Therefore it is natural to assume
that $K$ and $a_{\perp}\lambda$ are only functions of $l$ and $\tau$.
The only dimensionally correct expressions are then \cite{politi}
\begin{equation}
K \approx a_{\perp}\lambda \approx l^4/\tau.
\label{K2}
\end{equation}

By contrast, the shot noise cannot depend on surface diffusion.
Therefore, the diffusion length $l$ cannot enter, and
${\cal F}/a_{\perp}^2$ can only be a function of $a$ and $\tau$.
The only dimensionally
correct expression is then
\begin{equation}
{\cal F} \approx a_{\perp}^2 a^d /\tau = F (a_{\perp} a^d)^2
\label{F2}
\end{equation}.

Comparing (\ref{K2},\ref{F2}) with (\ref{K1},\ref{F1}) one finds that
\begin{equation}
\tilde l^4/\tilde t \approx l^4/\tau \qquad {\rm and} \qquad 
\tilde l^d/\tilde t \approx a^d/\tau .
\label{zwischenergebnis}
\end{equation}
This, finally, leads to the central result of this paper,
\begin{equation}
\tilde l/a \approx (l/a)^{4/(4-d)} \qquad {\rm and} \qquad
F a^d \tilde t = \tilde t/\tau \approx (l/a)^{4d/(4-d)} .
\label{result}
\end{equation}
Note in particular
that the layer coherence length $\tilde l \gg{}\l$.
With (\ref{l}) the exponent $\delta$ defined in (\ref{tilde_t}) is
\begin{equation}
\delta = \gamma \frac{4d}{4-d} ,
\label{delta}
\end{equation}
provided the cKPZ equation is the appropriate continuum equation for the
growth process.
At the upper critical dimensionality $d = d_c = 4$ the scales $\tilde l$ and 
$\tilde t$ depend 
exponentially on $l$, while for $d > d_c$ the oscillations persist
forever and the surface remains smooth.

\section{The adatom current and the shot noise reconsidered}
\label{current}

In this section (\ref{K2}) and (\ref{F2}) will be rederived without
using dimensional arguments. First we give a microscopic derivation of 
the nonlinear contribution to the adatom current (\ref{surface_current}) 
(see also \cite{advances}).

It was proposed by Villain \cite{villain91} that in growth processes
far from equilibrium, where local chemical potentials along the 
surface are ill defined, diffusion currents should be driven by gradients
in the growth-induced, nonequilibrium adatom density $n$,
\begin{equation}
{\bf j}= - D a_{\perp} a^d \ \nabla n  .
\label{villain_current}
\end{equation}
The factor $a^d a_{\perp}$ is the atomic volume and enters because
(\ref{cKPZ}) expresses volume rather than mass conservation.

On a singular surface the balance between deposition and capture
of adatoms at steps leads to a stationary adatom density $n = n_0$ of the
order \cite{villain92}
\begin{equation}
\label{n0}
n_0 \approx (F/D) l^2.
\end{equation}
On a vicinal surface
the adatom density is reduced due to the presence of additional steps;
however this effect is felt only if the miscut $m=|\nabla h|$ exceeds 
$a_{\perp}/l$,
in which case (\ref{n0}) is replaced by $n \approx (F/D) (a_{\perp}/m)^2$. 
In terms of a coarse grained description of the surface this implies that
the local adatom density depends on the local miscut or surface 
tilt. A useful interpolation formula which connects 
the regimes $m \ll{}a_{\perp}/l$ and $m \gg{}a_{\perp}/l$ is \cite{politi}
\begin{equation}
\label{n(m)}
n(\nabla h) = \frac{n_0}{1 + (l \nabla h/a_{\perp})^2} \approx (F/D) l^2 - 
(F/D) l^4 (\nabla h/a_{\perp})^2 + \dots
\end{equation}
Inserting the leading quadratic term of this gradient expansion into
(\ref{villain_current}), which is appropriate for describing
long wavelength fluctuations around the singular orientation, we obtain
\begin{equation}
\label{villaineq}
{\bf j} = \nabla \lambda (\nabla h)^2 
\end{equation}
with
$\lambda = F a^d l^4/a_{\perp}$, which agrees with the result (\ref{K2}) of
the previous section.

Now we rederive (\ref{F2}) for the shot noise, which can be written as
\begin{equation}
\tilde l^d/\tilde t = F a^{2d}
\label{F3}
\end{equation}
by inserting (\ref{F1}) and (\ref{tau}). This follows from 
considering the
number of particles deposited during time $\tilde t$ into area
$\tilde l^d$, $F \tilde t \ \tilde l^d
\pm{}(F \tilde t \ \tilde l^d )^{1/2}$. Each particle contributes a 
volume $a^{d}a_{\perp}$, so that the fluctuation of the film
thickness over the distance $\tilde l$ is 
\begin{equation}
w(\tilde t) \approx \sqrt{F \tilde t \ \tilde l^d} \  
                    a^{d}a_{\perp}/\tilde l^d .
\label{tilde_w}
\end{equation}
At $\tilde t$ this should be the thickness of about 
one atomic layer, $w(\tilde t) \approx a_{\perp}$, which results 
in (\ref{F3}).

\section{Competing mechanisms}
\label{competition}

Whereas for our computer simulation model the theoretical arguments
given in the preceding sections are perfectly appropriate, the
question arises how relevant these results are in general experimental
situations. It has been argued that generically one should expect
nonequilibrium contributions to the surface current which are driven
by a height difference \cite{villain91,wolf_villain,kps}.
To leading order in a gradient expansion one gets an adatom current
of the form
\begin{equation}
\label{ew}
{\bf j}=-\nu \nabla h .
\end{equation}
Eq.(\ref{cKPZ}) with such a current is known as the Edwards-Wilkinson (EW) 
equation \cite{EW}. 

Tilt induced nonequilibrium
surface currents originate from 
step edge barriers of Ehrlich-Schwoebel-type 
\cite{ehrlichschwoebel}, as well as
kick-out or
diffusion exchange processes at step edges.
Whereas the latter two lead to a downhill current stabilizing the
surface ($\nu > 0$), the former
generates an uphill current ($\nu < 0 $) and consequently an instability which
will not be considered in the following.

In the case of kick-out processes the coefficient $\nu$ cannot depend
on the diffusion length, because they are caused by deposition events 
in the immediate vicinity of a downward step. The only dimensionally
correct expression is therefore
\begin{equation}
\label{nu1}
\nu = a^2/\tau= F a^{d+2}.
\end{equation}
The corresponding current is proportional to the local step density $\nabla h$
and the deposition rate $F$. 
Such contributions will be absent in the simulations discussed in the next 
section.

In general, the adatom current will contain the terms (\ref{surface_current})
as well as (\ref{ew}). The latter one dominates 
the surface roughness on large scales.
Whether or not it influences the damping of the growth oscillations,
however, depends on the crossover time $t_{\lambda\nu}$ from cKPZ-
($\lambda$-dominated) at 
early to EW- ($\nu$-dominated) behaviour at late times. 
If the oscillations are damped out before the crossover takes place,
the $\lambda$-term determines the damping, hence the above result
applies. Let 
$\tilde t_{\lambda}$ and $\tilde t_{\nu}$ 
denote the damping times if only the $\lambda$- or the $\nu$-term were
present in the continuum equation of motion. Then (\ref{result}) and 
(\ref{delta}) hold if
\begin{equation}
\tilde t_{\lambda} \leq t_{\lambda\nu} .
\end{equation}
If, however, this is not the case, then $\tilde t_{\lambda}$ is replaced by
$\tilde t_{\nu}$, as long as no further
terms in the continuum description provide further time scales.

The crossover time $t_{\nu\lambda}$ is estimated in the following way:
First we calculate the typical height fluctuation $h_{\nu}(t)$ after
time $t$, if the $\lambda$-term would be absent. Similarly, $h_{\lambda}(t)$ 
is the fluctuation amplitude, if $\nu = 0$. Equating $h_\lambda$ and
$h_\nu$ then gives $t_{\lambda\nu}$.
By dimensional analysis one gets (see appendix)
\begin{equation}
\label{lambdanu}
{\cal F}t_{\lambda\nu} \approx \left(\frac{\lambda}{\cal F}\right)^{4/(d+2)}
\left(\frac{\cal F}{\nu}\right)^{(d+8)/(d+2)} \approx a_{\perp}^2 a^d \left(
\frac{l}{a}\right)^{16/(d+2)} ,
\end{equation}
where (\ref{tau}-\ref{F2}) and (\ref{nu1}) have been used to replace the 
parameters $\lambda$, ${\cal F}$ and $\nu$.

This has to be compared with (\ref{result}),
\begin{equation}
{\cal F} \tilde t_{\lambda} \approx a_{\perp}^2 a^d 
\left(\frac{l}{a}\right)^{4d/(4-d)}.
\label{t_lambda}
\end{equation}
For $d \leq 2$ the damping time $\tilde t_{\lambda}$ is smaller or of equal 
order of magnitude as the crossover time $t_{\lambda\nu}$. This implies
that kick-out processes at step edges, although leading to an EW-term in the
growth equation and hence modifying the later roughness, do not change our
results (\ref{result}) for the layer coherence length and the damping time.

However, if for example the sticking probability at an up step would be
much smaller than at a down step (e.g. due to a step decoration by surfactant
atoms \cite{markov}), one would expect a downhill current depending on $l$ rather than the
lattice constant $a$, i.e. with
\begin{equation}
\nu \approx l^2/\tau
\label{nu2}
\end{equation}
instead of (\ref{nu1}). In this case (\ref{lambdanu}) is replaced by
\begin{equation}
{\cal F} t_{\lambda\nu} \approx a_{\perp} a^d
\left(\frac{l}{a}\right)^{- 2d/(2+d)},
\end{equation}
which is never larger than ${\cal F} \tilde t_{\lambda}$. 
The damping time should then be given by (\ref{last})
\begin{equation}
F a^d \tilde t_{\nu} \approx \left(\frac{l}{a}\right)^{2d/(2-d)}.
\end{equation}

\section{Numerical Results}
\label{numerics}

We carried out simulations with a simplified model:
Atoms are deposited onto a one-dimensional surface with a 
deposition rate $F$ and
diffuse with a diffusion constant $D$ until they meet other
adatoms or a cluster of adatoms. If the encountered cluster consists
of $i^*$ or more atoms, the atom is incorporated into this cluster, and
a stable, immobile island is formed. 
Clusters of size $i^*$ or less are allowed to decay. There are no
barriers for interlayer transport (Ehrlich-Schwoebel barriers
  \cite{ehrlichschwoebel})
and no overhangs or holes are allowed (solid-on-solid condition).
In the simulations we use $a$ and $a_{\perp}$ as length and height unit, i.e.
$a=a_{\perp}=1$.

\begin{figure}[htb]
\centerline{\psfig{figure=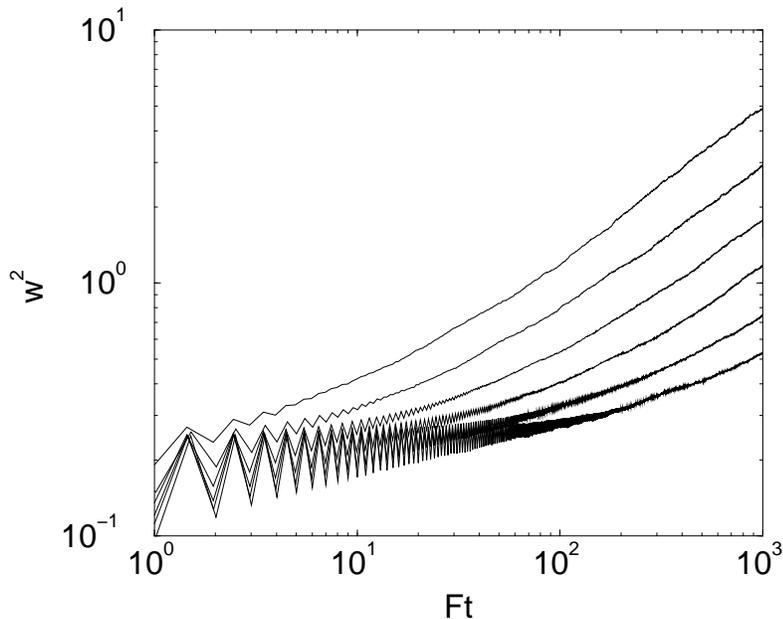,height=10.0cm,angle=270}}
\caption{Squared surface width at integer and half-integer times (in
units of the layer completion time $\tau=1/F$) for different
  values of the parameter $D/F=10^5\dots10^{10}$ (from top to bottom).}
\label{wi1}
\end{figure}

\begin{figure}[htb]
\centerline{\psfig{figure=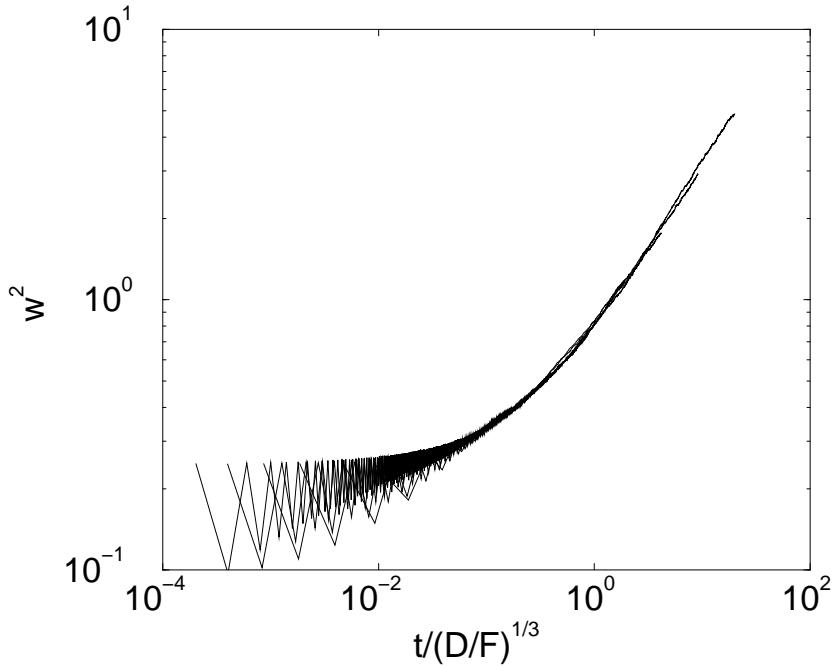,height=10.0cm,angle=270}}
\caption{Curves from fig.~\ref{wi1}, with time scaled by $(D/F)^{1/3}$.}
\label{wi1_collapse}
\end{figure}

The squared surface width $w^2=\langle h^2\rangle-\langle h \rangle^2$ 
as a function of time for various $D/F$
and $i^*=1$ is shown in fig.~\ref{wi1} ($\langle\dots\rangle$ denotes 
the ensemble and spatial average).
For a given $D/F$, the oscillations in the surface width persist up to
a coverage $F\tilde t$, which increases with $D/F$.
Beyond $F\tilde t$, the crossover to kinetic
roughening is observed, where $w^2$ approaches a power law $t^{2\beta}$ 
with the cKPZ
prediction of $\beta = 1/3$ in one dimension \cite{villain91,dasSarma,CKPZ}.
Rescaling $Ft$ by $F\tilde t = (D/F)^{1/3}$ in fig.\ref{wi1_collapse}
we find an excellent collapse of 
the crossover regions for all curves of fig.\ref{wi1}. This means that 
$\delta=1/3$ within numerical accuracy, 
in agreement with (\ref{delta}) and $\gamma=1/4$ 
\cite{pimpinelli}.

\begin{figure}[htb]
\centerline{\psfig{figure=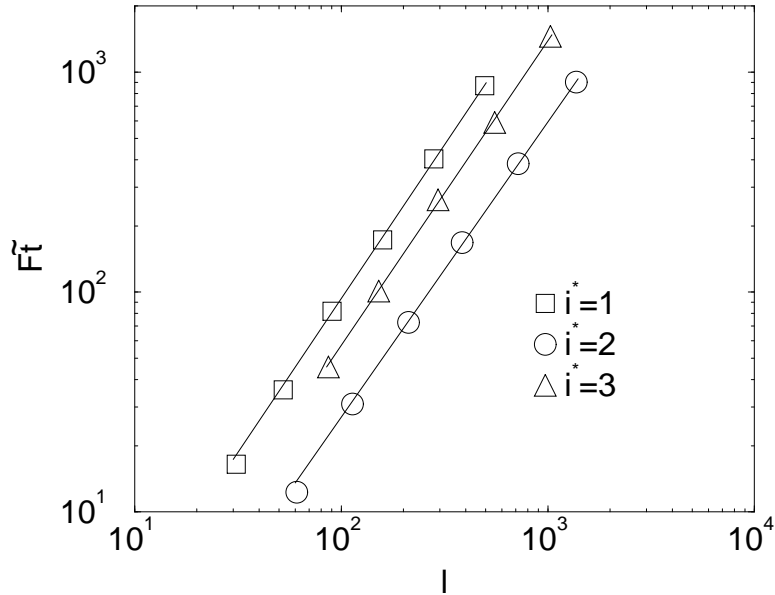,height=10.0cm,angle=270}}
\caption{Coverage $F\tilde t$, at which the surface width reaches a given
value (see text), as a function of the diffusion length $l$
  for different values $i^{*}$.   The straight lines are fits to the last four 
  data points in each set of data. Their slopes are 
  $1.39\pm{}0.09$, $1.34\pm{}0.09$ and $1.38\pm{}0.09$ for $i^*=1, 2, 3$,
  respectively.
}
\label{tcl}
\end{figure}

We measured the damping time for different values of $i^*$ by determining 
the coverage $F\tilde t$, where $w=0.71$, 0.57 or 0.65 for $i^*=1$, 2, 3, 
respectively. It depends on $l$ measured as the system size divided by the
total number of nucleation events which occurred in one layer. Fig.\ref{tcl}
shows that
$F\tilde t\sim l^{4/3}$, independent of the values 
$i^*=1, 2, 3$, in agreement with the theoretical result (\ref{result}).

\begin{figure}[htb]
\centerline{\psfig{figure=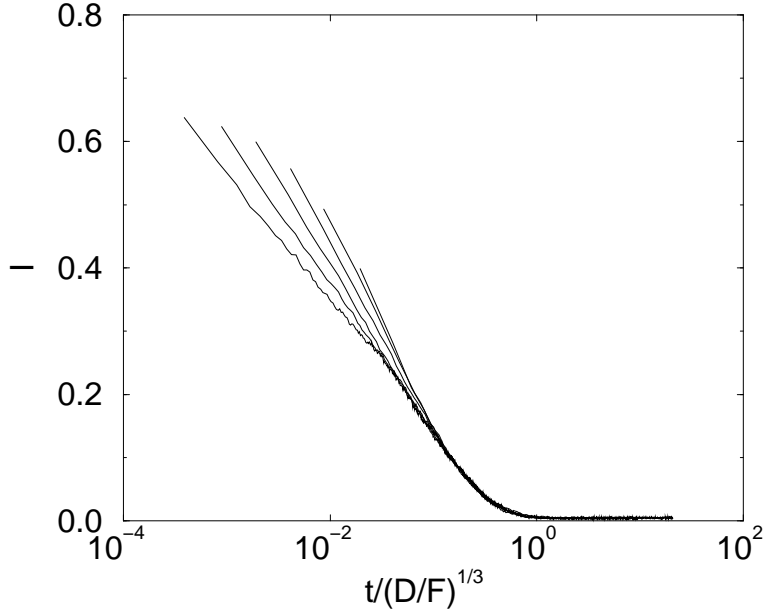,height=10.0cm,angle=270}}
\caption{Maxima of the kinematic intensity for 
  $D/F=10^5\dots 10^{10}$ (right to left), and $i^*=1$. 
  Time is scaled by $(D/F)^{1/3}$.}
\label{ki1_collapse}
\end{figure}

In order to check that the damping time and the layer coherence length are
the appropriate scales also for other quantities showing oscillations during
the layer-by-layer growth we investigated the
Bragg intensity or kinematic intensity 
$I=\langle(n_{even}-n_{odd})^2\rangle$
where $n_{even}$ ($n_{odd}$) denotes the number of surface
sites in even (odd) layers. It is related to the RHEED intensity in
experiments. Fig.~\ref{ki1_collapse} shows $I$ at integer times,
rescaled in the same way as in fig.\ref{wi1_collapse}.
Again, we find that the number of observable oscillations varies with
the growth conditions as described by (\ref{result}).

\begin{figure}[htb]
\centerline{\psfig{figure=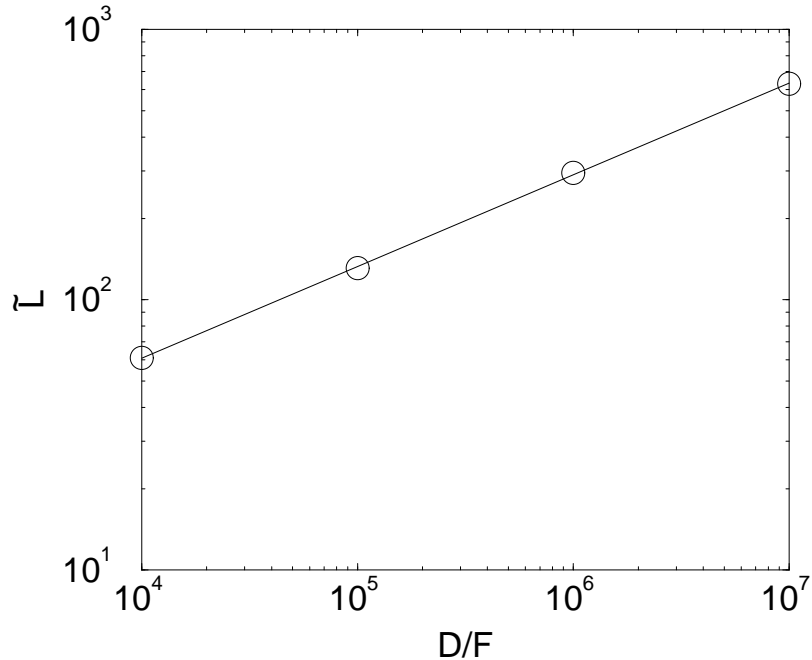,height=10.0cm,angle=270}}
\caption{$\tilde L$ as a function of $D/F$.
  The fit has a slope of $0.339\pm{}0.006$.}
\label{ltilde}
\end{figure}

Finally, we carried out a finite size analysis to measure the layer
coherence length $\tilde l$
explicitly. As mentioned in section \ref{introduction}, the surface does not
roughen when the system size $L$ is smaller than $\tilde l$. 
Then, the oscillations of the surface width $w$ persist forever.
After a transient time, the amplitude
of the oscillations becomes stationary. We take the variance of the surface
width $w(t)$ during the layer completion time $\tau=1/F$,
\begin{equation}
A(t)^2=\langle w^2\rangle_{[t,t+\tau]} - \langle w\rangle^2_{[t,t+\tau]},
\end{equation}
as a measure of the squared amplitude of the oscillations. 
$\langle \dots \rangle_{[t,t+\tau]}$ means the time average over the 
interval $[t,t+\tau]$. If this variance
becomes equal to the ensemble fluctuations of $w$ at fixed time, no 
oscillations can be observed.
We find that $A(t)$ approaches a stationary value which decreases with
increasing system size. For system sizes larger than a certain value $\tilde L$
$A(t)$ is equal to the statistical fluctuations of $w$ itself. 
This means that in a system of
size $L>\tilde L$ the oscillations can die out
(or rather cannot be distinguished from noise anymore for long times).
Therefore,
$\tilde L$ can be identified with $\tilde l$. 
According to (\ref{result}) and (\ref{l}) with $\gamma = 1/4$ 
in one dimension for $i^*=1$ \cite{pimpinelli}, one
expects $\tilde L\sim (D/F)^{1/3}$.
Indeed, the simulation results shown in fig.~\ref{ltilde} are in
excellent agreement with the theoretical prediction.

\section{Acknowledgements}
Useful discussions with M.~Schroeder, P.~\v{S}milauer and J.~Villain
are gratefully acknowledged.
We acknowledge support by DFG within SFB 166 {\em Strukturelle und magnetische
Phasen\"uberg\"ange in \"Ubergangsmetall-Legierungen und Verbindungen} (D.E.W.)
and SFB 237 {\em Unordnung und grosse Fluktuationen} (J.K.).

\section*{Appendix}
\label{appendix}

We consider the following continuum equation :
\begin{equation}
\label{all}
\frac{\partial h}{\partial t} = \nu\nabla^2 h - K(\nabla^2)^2 h
-\lambda\nabla^2(\nabla h)^2 + \eta + \eta_c
\end{equation}
where $\eta$ and $\eta_c$ are gaussian distributed random forces with
zero mean and second moment
\begin{eqnarray}
  \langle\eta({\bf x},t)\eta({\bf x^{'}},t^{'})\rangle &=& {\cal F}~
  \delta^d({\bf x}-{\bf x^{'}})~\delta(t-t^{'}),\\
  \langle\eta_c({\bf x},t)\eta_c({\bf x^{'}},t^{'})\rangle& =& -{\cal D}~
  \nabla^2~  \delta^d({\bf x}-{\bf x^{'}})~\delta(t-t^{'}),
\end{eqnarray}
which describe the shot noise and the diffusion noise \cite{sgg},
respectively. 
Dimensional arguments as in section \ref{exponent} together with 
(\ref{n0}) and (\ref{F2})lead to the well known expression
\begin{equation}
{\cal D} \approx {\cal F} l^2 \approx D n_0 \  (a^d a_{\perp})^2
\label{D}
\end{equation}
for the correlator of the conserved noise \cite{CKPZ}. This implies that
the conserved noise dominates the fluctuations only on distances shorter
than the typical diffusion length $l$ \cite{moser_wolf}. As we are
dealing with larger length scales, the conserved noise may be neglected
in the following.

The physical dimensions of the remaining parameters in (\ref{all}) are
\begin{equation}
\label{dim}
  [\nu]=x^2t^{-1}, [K]=x^4t^{-1}, [\lambda]=x^4t^{-1}h^{-1}, 
 [{\cal F}]=x^dt^{-1}h^2.
\end{equation}
Comparing those of $\nu$ and ${\cal F}$ one gets
\begin{equation}
h_{\nu}(t) = ({\cal F}/\nu)^{d/4} ({\cal F}t)^{(2-d)/4} ,
\label{hnu}
\end{equation}
comparing those of $\lambda$ and ${\cal F}$ one gets \cite{amar_family}
\begin{equation}
h_{\lambda}(t) = ({\cal F}/\lambda)^{d/(8+d)} ({\cal F}t)^{(4-d)/(8+d)},
\label{hlambda}
\end{equation}
and finally comparing those of $K$ and ${\cal F}$ one gets
\begin{equation}
\label{hk}
h_K(t) = ({\cal F}/K)^{d/8} ({\cal F}t)^{(4-d)/8}.
\end{equation}

The dimensional analysis of the linear equations, leading to
(\ref{hnu}) and (\ref{hk}), already gives the right scaling behaviour
of $h$ as function of $t$,
due to the non-renormalisation of the parameters $\nu$, $K$ and ${\cal F}$ 
\cite{advances}.

Setting 
$h_\lambda(t_{\lambda\nu}) = h_\nu(t_{\lambda\nu})$ gives the
crossover time $t_{\lambda\nu}$ in 
(\ref{lambdanu}).
In the same fashion, by setting $h_K(t_{K\lambda}) = h_\lambda(t_{K\lambda})$ 
one gets
\begin{equation}
\label{tklambda}
{\cal F}t_{K\lambda} = \left(\frac{K}{\cal F}\right)^{(8+d)/(4-d)}
\left(\frac{\cal F}{\lambda}\right)^{8/(4-d)} = a_{\perp} a^d \left(\frac{l}{a}\right)^{4d/(4-d)} 
\end{equation}
for the crossover time from $K-$ to $\lambda-$dominated roughening.
In the last equality eqns. (\ref{K2}) and (\ref{F2}) have been used.
Then the crossover time agrees with the expression (\ref{t_lambda}), 
consistent with the fact, that the $K$-term and the $\lambda$-term give the
same result.

Finally, one can ask for the typical times, where $h_K$, $h_{\lambda}$
or $h_{\nu}$ become $\simeq 1$,
i.e. the times which can be interpreted as the damping times, if only
the corresponding term is present:
\begin{equation}
{\cal F}\tilde t_K = a_{\perp}^2\left(\frac{K a_{\perp}^2}{\cal F}\right)^{d/(4-d)}, 
{\cal F}\tilde t_\lambda = a_{\perp}^2\left(\frac{\lambda a_{\perp}^3}{\cal F}\right)^{d/(4-d)},
{\cal F}\tilde t_\nu = a_{\perp}^2\left(\frac{\nu a_{\perp}^2}{\cal F}\right)^{d/(2-d)}. 
\label{last}
\end{equation}


\begin{thebibliography}{99}


\bibitem{VP95} J.~Villain, A.~Pimpinelli: {\em Physique de la Croissance 
Cristalline} (\'Editions Eyrolles et CEA, 1995)
\bibitem{bauer} E.~Bauer, Z. Kristallogr. {\bf 110}, 372 (1958)
\bibitem{evans} 
H.C.~Kang and J.W.~Evans, Surf.~Sci. {\bf 271}, 321 (1992);
M.C.~Bartelt and J.W.~Evans, Phys.~Rev.~Lett. {\bf 75}, 4250 (1995).
\bibitem{villain91} J. Villain, J. Phys. France I {\bf 1}, 19 (1991).
\bibitem{siegert}  M.~Siegert and M.~Plischke, Phys.~Rev.~Lett.~{\bf 68},
2035 (1992); ibid. {\bf 73}, 
1517 (1994); P.~\v{S}milauer and D.D.~Vvedensky, 
Phys.~Rev.~B {\bf 52}, 14263 (1995).
\bibitem{experiments}
K.~Th\"urmer, R.~Koch, M.~Weber and K.H.~Rieder, Phys.~Rev.~Lett.{\bf 75},
1767 (1995);
J.A. Stroscio, D.T. Pierce, M. Stiles, A. Zangwill and L.M. Sander,
Phys. Rev. Lett. {\bf 75}, 4246 (1995);
J.E. Van Nostrand, S.J. Chey, M.-A. Hasan, D.G. Cahill and J.E. Greene,
Phys. Rev. Lett. {\bf 74}, 1127 (1995).
\bibitem{zinsmeister} G.~Zinsmeister, Thin Solid Films {\bf 2}, 497 (1968);
ibid. {\bf 4}, 363 (1969); ibid. {\bf 7}, 51 (1971).
\bibitem{stoyanov} S.~Stoyanov and D.~Kashchiev in {\em Current Topics
    in Material Science}, edited by E.~Kaldis (North-Holland,
  Amsterdam, 1981), Vol.~7, pp. 69 - 141.
\bibitem{venables} J.~A.~Venables, G.~D.~Spiller and M.~Hannbrucken,
  Rep.~Prog.~Phys.~{\bf 47}, 300 (1984).
\bibitem{villain92} J. Villain, A. Pimpinelli and D. Wolf,
Comments Cond. Mat. Phys. {\bf 16}, 1 (1992).
\bibitem{pimpinelli} A.~Pimpinelli, J.~Villain, D.~E.~Wolf,
  Phys.~Rev.~Lett.~{\bf 69}, 985 (1992).
\bibitem{wolf} D.~E.~Wolf, in: {\em Scale Invariance, Interfaces, and 
Non-Equilibrium Dynamics}, eds. A. McKane, et al. (Plenum, New York, 1995) 
pp. 215 - 248.
\bibitem{pimpinelli96} P.~Jensen, H.~Larralde and A.~Pimpinelli, 
{\tt cond-mat/9610001}
\bibitem{somfai} E.~Somfai, J.~Kert\'esz and D.~E.~Wolf,
  J.~Phys.~I~France~{\bf 6}, 393 (1996).
\bibitem{brendel}L.~Brendel: {\em Fluktuationsschw\"achung in 
Wachstumsmodellen f\"ur Molekularstrahlepitaxie}, Diplomarbeit, 
Gerhard-Mercator-Univ. Duisburg, 1994.
\bibitem{family_vicsek} F.~Family and T.~Vicsek (eds.): 
{\em Dynamics of Fractal Surfaces} (World Scientific, Singapore 1991).
\bibitem{wolf_villain} D.~E.~Wolf and J.~Villain, Europhys. Lett. {\bf 13}, 389
(1990).
\bibitem{dasSarma} Z.-W. Lai and S. Das Sarma, Phys. Rev. Lett. {\bf 66},
2348 (1991).
\bibitem{CKPZ} L.~-~H.~Tang and T.~Nattermann, Phys.~Rev.~Lett.~{\bf
    66}, 2899 (1991).
\bibitem{politi} P. Politi and J. Villain, Phys.~Rev.~B~{\bf 54}, 5114
  (1996).
\bibitem{advances} J. Krug, Adv. Phys. (in press).
\bibitem{kps} J. Krug, M. Plischke and M. Siegert, 
Phys. Rev. Lett. {\bf 70}, 3271 (1993).
\bibitem{EW} S.F. Edwards and D.R. Wilkinson, 
Proc. Roy. Soc. London A {\bf 381}, 17 (1982).
\bibitem{ehrlichschwoebel} G.~Ehrlich and F.~G.~Hudda,
  J.~Chem.~Phys.~{\bf 44}, 1039 (1966); R.~L.~Schwoebel and
  E.~J.~Shipsey, J.~Appl.~Phys.~{\bf 37}, 3682 (1966).
\bibitem{markov} I.~Markov, Phys. Rev. B {\bf 50}, 11271 (1994)
\bibitem{sgg}
T. Sun, H. Guo and M. Grant, Phys. Rev. A {\bf 40}, 6763 (1989).
\bibitem{moser_wolf} K.~Moser and D.~E.~Wolf, in: {\em Surface Disordering:
Growth, Roughening, and Phase Transitions}, eds. R.~Jullien, J.~Kert\'esz,
P.~Meakin, and D.~E.~Wolf (Nova Science, Commack, 1992) p. 21
\bibitem{amar_family} J.~Amar and F.~Family, Phys.~Rev.~A {\bf 45}, 5378 (1992)




\end{thebibliography}
\end{document}